\documentstyle[prl,aps]{revtex}

\begin{document}

\title{Lyapunov exponents, entropy production and decoherence}
\author{Arjendu K. Pattanayak}
\address{Department of Physics, Rice University, Houston, TX 77251-1892}
\maketitle
\widetext
\begin{abstract}
{We establish that the entropy production rate of a classically chaotic 
Hamiltonian system coupled to the environment settles, after a transient, 
to a meta-stable value given by the sum of positive generalized Lyapunov 
exponents. A meta-stable steady state is generated in this process.  
This behavior also occurs in quantum systems close to the classical limit
where it leads to the restoration of quantum-classical correspondence in 
chaotic systems coupled to the environment. }
\end{abstract}

The Lyapunov exponents of a chaotic classical Hamiltonian system characterize 
the rate of exponential divergence of neighbouring phase-space trajectories 
thus defining the microscopic time-scales. 
The question\cite{gaspard} of their role in the macroscopic behavior of 
the system is hence at the foundation of statistical mechanics. It is
also of deep interest to understand how the Lyapunov exponents govern the 
behavior of the quantum counterparts of these chaotic systems\cite{97_1}. 
For instance, it is predicted\cite{berman} that quantum effects comparable 
to the classical behavior should be measurable on a time-scale 
$t_B \approx \frac{1}{\lambda}\ln(\frac{1}{\hbar})$. Here $\hbar$ is Planck's 
constant scaled by a characteristic action for the system and $\lambda$ 
is the largest Lyapunov exponent of the system. This `break-time' can
be $\approx 20$ years for astronomical systems\cite{zp} and is clearly 
incompatible with observations. It is argued\cite{zp} that this problem 
may be resolved by considering the decoherence resulting from including 
environmental effects on the system evolution. Since this can be understood 
as a dynamical coarse-graining, and statistical properties of the system are 
typically obtained through a static coarse-graining, a diagnostic of common 
primary interest for both issues is the behavior of the coarse-grained 
entropy of the system. 
The Gibbs entropy $S_G ={\rm Tr}[\rho\ln\rho]$ and the linear or Renyi entropy 
$S_2 =\ln({\rm Tr}[\rho^2])$ are both useful\cite{entropy} in this regard. 
Here $\rho$ is the classical phase-space probability distribution $\rho^C(x)$ 
or the quantum quasi-probability Wigner distribution $\rho^W(x)$ as 
appropriate and Tr denotes the integration over all phase-space variables 
$x =(q,p)$. These entropies measure the information in the probability 
distribution and remain constant for Hamiltonian evolution without 
coarse-graining. A change in the coarse-grained entropy corresponds to 
information being lost at the finest scales. Since in a chaotic system 
$\rho$ acquires structure at small scales exponentially rapidly\cite{97_1}, 
$\dot S_2$ typically grows exponentially rapidly in magnitude initially for 
such classical systems, as also for their quantum analogs close to the 
classical limit\cite{97_2}. 
Numerical studies with static coarse-graining\cite{static} and the
analysis of the behavior of a Gaussian wave-packet in the upside-down 
harmonic oscillator suggest that for weakly coupled systems $\dot S_G$ 
settles to a meta-stable `constant' after this transient. This meta-stable 
value is conjectured\cite{zp} to be the sum $\sum_i\lambda_i^+$ over the 
positive Lyapunov exponents of the classical system, independent of the 
details of the original distribution. This conjecture has not yet been
verified. Preliminary analyses\cite{miller} indicate that systems may
become meta-stable although the quantitative conjecture does not hold in
general. In this paper, we consider a variation of this conjecture and 
are able to analytically establish it for the {\em general} case of an 
arbitrary distribution evolving in a chaotic system. 
We do so by considering the dynamics of the quantity $\chi(t)$ which measures 
the degree of structure in a phase-space distribution, and is directly 
proportional to the entropy production rate. We demonstrate that under 
reasonable approximations the entropy production rate $\dot{S_2}$ indeed 
settles after an exponential transient to 
$\dot{S_2} \approx -\sum_i \Lambda_{2,i}^+$. The $\Lambda_{2,i}$, 
as defined below, are generalized Lyapunov exponents that depend upon the 
initial distribution in general. We further show that this is a robust 
result: Although for sufficiently weakly coupled systems $\dot{S_2}$ 
{\em increases} in magnitude to a constant, for a stronger coupling the 
magnitude of $\dot{S_2}$ starts high and {\em decreases} to the same constant. 
The meta-stable rate is hence independent of the environmental strength.
We also generalize the idea that for such systems $\rho$ can evolve to a 
meta-stable state\cite{zp,bag} and show that for such states the quantum 
corrections for classically chaotic systems remain small for all times, 
indeed resolving the issue of the anomalously short `break-time'. 
We thus obtain a condition for the 
restoration of correspondence through decoherence. Finally, we present 
numerical results verifying these analytic results for the particular case 
of the quantum Cat Map and its classical limit, a uniformly hyperbolic 
K-system satisfying the assumptions in our arguments\cite{quantum_cat}.

The evolution of $\rho^W$ under the potential $V(q)$ and coupled to an 
external environment\cite{decoh,zp} is
\begin{equation}
{\partial \rho^W\over\partial t}=\{H,\rho^W\} 
+\sum_{n \geq
1}\frac{\hbar^{2n}(-1)^n}{2^{2n} (2n +1)!}
\frac{\partial^{2n+1} V(q)}{\partial
q^{2n+1}}\; \frac{\partial^{2n+1} \rho^W}{\partial p^{2n+1}}
+ D \nabla^2\rho^W.
\label{wigner}
\end{equation}
The first term on the right is the Poisson bracket, generating the classical 
evolution for $\rho^W$; the terms in $\hbar$ add the quantal evolution.
The environmental effects are contained in the $\nabla^2$ term; for simplicity 
we have assumed coupling to all the phase-space variables, although the results 
can be easily generalized. The parameter $D$ depends strongly upon the form 
of the coupling and the spectrum of the environment\cite{decoh}. The entropy 
production rate is then 
$\dot{S_2} = 2D\frac{{\rm Tr}[\rho^W \nabla^2\rho^W]}{{\rm Tr}[(\rho^W)^2]} = 
-2D \frac{{\rm Tr}[|\nabla \rho^W|^2]}{{\rm Tr}[(\rho^W)^2]} 
\equiv -2D\chi^2$
where the second equality results from an integration by parts. The 
global quantity $\chi^2(t)$ corresponds to the mean-square radius of 
the Fourier expansion of $\rho$; it hence measures the structure in the 
distribution\cite{97_1} and also governs the entropy production rate. 
To understand the behavior of $S_2$ first consider the classical Hamiltonian 
evolution, that is, only the Poisson bracket terms in Eq.~(\ref{wigner}). 
In this case, the point dynamics satisfy $\dot {x} = f(x)$ with the 
accompanying induced flow $x_t = \Phi^tx_0$ and\cite{97_1}
the phase-space gradient along a trajectory is governed by
${\nabla}\rho(x_t)=-{\cal M}(t,x_0){\nabla}\rho(x_0)$, 
where the fundamental matrix ${\cal M}$ is given by\cite{gaspard} the 
time-ordered series ${\cal M}(t,x_0) = {\cal T} \exp \int_0^t 
\frac{\partial f (\Phi^{\tau}x_0)}{\partial x}d\tau$. This defines a 
real symmetric matrix ${\cal M}^T\cdot{\cal M}$ (the transpose denoted
by ${\cal M}^T$) which can be diagonalized\cite{gaspard} as 
$M^T(t,x_0)\cdot M (t,x_0) =
\sum_{i} u_i(t,x_0)\sigma_i(t,x_0)u_i^T(t,x_0)$,
whence the $u_i$ constitute a local orthonormal tangent space for the flow.
The local Lyapunov exponents are then given by
$\lambda_i(x_0) = \lim_{t\to\infty} \frac{1}{2t}\ln\sigma_i(t,x_0).$
We may now define the global averages 
\begin{equation}\label{lambda}
\Lambda_{2,i}(\rho,t) =
\frac{1}{2t}
\ln \bigg[ \frac{{\rm Tr}[|\nabla_i\rho(x_0)|^2
\sigma_i(t,x_0)]}{{\rm Tr}[\nabla_i\rho(x_0)|^2}\bigg ]
= \frac{1}{2t}
\ln \bigg[ \frac{{\rm Tr}[|\nabla_i\rho(x_t)|^2]}
{{\rm Tr}[\nabla_i\rho(x_0)|^2}\bigg ]
\end{equation}
and their limits $\Lambda_{2,i}(\rho) = \lim_{t\to\infty}\Lambda_{2,i}(\rho,t)$.
The last equality in Eq.~(\ref{lambda}) comes from the orthonormality 
of the $u_i$ and the dependence of $\nabla\rho(x_t)$ on ${\cal M}$.
These $\Lambda$ are $\rho$ dependent versions of the usual generalized 
Lyapunov exponents of second order\cite{schlogl} which may be recovered by 
replacing $|\nabla_i \rho(x_0)|^2$ by the natural invariant measure $\rho_0(x)$ 
of the dynamics. For linear systems $\sigma_i(t,x_0) = \sigma_i(t)$ and the 
definitions are independent of $\rho$. For ergodic dynamics 
$\Lambda_{2,i}(\rho,t) = \Lambda_{2,i}(\rho)$ for all time. However,
numerical evidence suggests that this equality is valid within small errors 
for `typical' initial densities $\rho$ for other Hamiltonian flows as well. 
We can always write $\Lambda_{2,i}(\rho,t) = \Lambda_{2,i}(\rho) +\xi_i/t$ 
where $\xi$ fluctuates in general and $\xi/t$ vanishes with increasing $t$. 
The analysis so far has been exact; hereafter we make the approximation of 
neglecting the $\xi$ term. 
We now decompose the entropy production rate into the contribution 
from the different stability subspaces as $\dot{S_2} = 
-2D\chi^2 = -2D\sum_i\chi_i^2$ where 
$\chi_i^2 = {\rm Tr}[|\nabla_i\rho(x_t)|^2]/{\rm Tr}[\rho^2(x_t)]
\equiv \langle k_i^2 \rangle_{\rho^2_k}$ is the $i$ component of $\chi^2$.
The preceeding enables us to write 
\begin{equation}
\frac{d \chi^2_i}{dt}  
\approx 2\chi^2_i\Lambda_{2,i}
\label{chaos}
\end{equation} 
or equivalently $\dot{S_2} \approx -2D\sum_i\chi^2_i(0)e^{2\Lambda_{2,i}t}$ 
where the $\rho$-dependence of $\Lambda_{2,i}$ is left implicit hereafter. 
This shows the initial behavior of $\dot{S_2}$; it is dominated by the largest 
positive generalized exponent $\Lambda^+_{2,1}$, thus recovering an earlier 
approximate result\cite{97_2}.
After this initial stage, the structure reaches the finest scales and
hence the $D$-dependent diffusion term becomes important. To include
the impact of this we first consider purely diffusive behavior such that 
${\partial_t\rho} = D \nabla^2\rho$. Since the Laplacian is independent of the 
coordinate system, this may be decomposed into the same subspaces as above 
as $\partial_t \rho = D\sum_i\nabla^2_i\rho$ with the solution being a 
product of the distributions as $\rho = \Pi_i\rho_i(x_i)$. 
Each direction has the standard solution written in terms of the
Fourier components as $\rho_{k_i} =\rho_{k_i}\exp(-Dk^2_it)$ whence
the individual
$\chi_i^2(t) =\langle k_i^2 \rangle_{\rho^2_k} =
\sum_k k_i^2 |\rho_k(0)|^2\exp(-2Dk^2t)/( 
{\sum_k|\rho_k(0)|^2\exp(-2Dk^2t)})$ have the time-dependence
\begin{equation}
\frac{d \chi^2_i}{dt} = -2D [\langle k_i^4 \rangle_{\rho^2_k}
- \langle k_i^2 \rangle^2_{\rho^2_k}]
\approx -4D\langle k_i^2 \rangle^2_{\rho^2_k} = -4D\chi_i^4.
\label{diff}
\end{equation}
The approximation in Eq.~(\ref{diff}) is a mean-field one, 
valid for the usual Gaussian solution to the diffusion equation for example. 
The various $\chi_i^2$ therefore behave as follows: An initial exponential 
transient entirely kills these quantities in the stable directions 
(corresponding to the negative Lyapunov exponents) and since this is 
enhanced by diffusive effects, these directions have a negligible role 
thereafter. In the unstable directions, however, the initial exponential 
growth is balanced by the diffusion. If we now explicitly set $d\chi_i^2/dt$ 
equal to the sum of the chaotic [Eq.~\ref{chaos}] and diffusive 
[Eq.~\ref{diff}] terms, we get
$\frac{d \chi^2_i}{dt} = 2\Lambda^+_{2,i}\chi^2_i - 4D\chi^4_i$
for the unstable directions. This has the stationary solution 
$\chi^{2*}_i = \Lambda^+_{2,i}/2D$ yielding the stationary entropy 
production rate 
$\dot{S_2}^* = -2D\chi^{2*} = -2D\sum_i\chi^{2*}_i = -\sum_i\Lambda^+_{2,i}$
with the sum over the positive exponents only, as just argued. 
Thus, within the approximations as above, we have shown that the entropy 
production rate for a chaotic system weakly coupled to the environment settles 
after an initial exponential transient to a meta-stable value as above, 
independent of the precise magnitude of the environmental 
effects\cite{footnote_n}. This solution is stable under small pertubations; 
thus, the exact dynamics arguably leave it unaltered. We now consider 
possible constraints on this result. First, note that the transition from 
the exponential to the linear regimes happens at
$t^*_i \approx (1/2\Lambda^+_{2,i})\ln [\Lambda^+_{2,i}/{2D\chi^2_i(0)}]$ 
[the shortest scale is approximated by the largest exponent $\Lambda^+_{2,1} 
\equiv \lambda$ and the full $\chi^2$ and we shall use those hereafter]. 
Since $t^* =0$ for $\beta \equiv 2\lambda^{-1}D\chi^2(0)=1$, for $\beta
\gg 1$, the initial behavior of $\chi^2_i$ is not the exponential transient 
but is given by Eq.~(\ref{diff}) instead. For this case, the initial 
diffusive effects are balanced by the chaos such that an initially large 
$\dot{S_2}$ decreases to a constant. 
Second, the difference between the initial entropy $S_2(0)$ and the final 
entropy $S_2(\infty)$ is typically finite. Thus, the system may not have 
`enough initial entropy' to evolve to the constant entropy rate as above. 
A rough estimate (and numerical results as described below) indicate that 
we need $S_2(0) - S_2(\infty) > 3$ for the linear saturation behavior to 
emerge\cite{expl}. 
Finally, in contrast to this behavior for chaotic systems, a similar analysis 
for non-chaotic systems yields an initial non-exponential decay of $S_2$ 
after which the competition from the diffusive term implies a maximum 
($\propto D^{-1}$) for $\dot{S_2}$, and a rapid decrease thereafter. As a 
result of these dynamics, appropriate initial distributions $\rho$ for
chaotic systems settle to a meta-stable state with $\chi^2_i$ as above. 
This is a remarkable result, generalizing earlier arguments\cite{zp,bag} for 
a `steady-state Gaussian width'. Further, this enables us to bound the 
quantum corrections in Eq.~(\ref{wigner}). Consider the first quantum 
term in Eq.~(\ref{wigner}), which scales as $\hbar^2\chi^3$. In the absence 
of the $D$-dependent terms, $\chi$ grows exponentially rapidly 
as $\exp(\lambda t)$, thus leading to a quantum-classical `break-time' 
$t_b \approx \frac{1}{\lambda}\ln(\frac{1}{\hbar})$ after which the 
quantum `correction' is comparable to the classical evolution and a 
classical description of the dynamics is invalid. This logarithmic 
dependence of $t_b$ on $\hbar$ is extremely weak and is a 
point of debate in the analysis of quantum chaotic systems\cite{zp}. 
However, environmental effects saturate $\chi$ and hence the first quantum 
term at $\zeta \approx \hbar^2 (\lambda/2D)^{3/2}$ such that this indeed is 
a `correction' of ${\cal O}(\hbar^2)$ to the classical evolution and 
there is no `break-time'. The condition\cite{kosl} for restoration 
of quantum-classical correspondence for chaotic systems may be summarized 
as $\zeta \ll 1$ (note that the other quantum terms are higher order in 
$\zeta$). Physically, quantum effects are important at the smallest 
scales $\approx \hbar$ of phase-space. Classically chaotic evolution 
increases the support of a distribution at the finest scales exponentially 
rapidly, thus enhancing quantum effects. However, noise or coarse-graining 
washes out the details of the fine-scale structure, thus restoring 
quantum-classical correspondence\cite{decoh-corresp}. For systems where
the above inequality is violated, however, a classical or semiclassical 
analysis breaks down and we must use the exact quantum evolution.

We have numerically verified the theory using the equivalent of 
Eq.~(\ref{wigner}) for the classical and quantum Cat Map\cite{quantum_cat}. 
As this system is linear and ergodic, the approximations above hold 
exactly. Some of the data obtained are shown in Figs.~(1--3). First, with 
a scaled Planck's constant $2\pi\hbar = 10^{-5}$, we expect\cite{quantum_cat} 
essentially classical behavior, even with no added noise. This is indeed 
verified as in Fig.~(1) where the classical data for the same initial 
conditions and $D$ exactly overlie the quantum results. For an initially 
sharply localized Gaussian state, corresponding to high initial entropy, 
when $D$ is small we see initial exponential entropy production\cite{97_2}; 
this saturates to linear behavior with the predicted slope. The initial 
exponential behavior is swamped as $D$ is increased, as argued previously. 
For a second initial state which is spread out and has lower initial entropy, 
we see similar stable behavior -- although $\dot{S_2}$ now saturates earlier, 
as discussed above -- confirming that this meta-stability is essentially 
independent of initial conditions and of $D$ for near-classical systems. 
Fig.~(2) shows the quantum-to-classical transition by comparing various 
runs for a single sharply localized initial condition at $2\pi\hbar= 10^{-3}$ 
with their classical equivalents. We see that for very small $D$ (hence 
large $\zeta$) there is significant quantal deviation from the classical 
behavior. However, as $D$ is increased (and $\zeta$ correspondingly decreases 
below $1$), the quantal $\dot{S_2}$ approaches the classical linear saturation 
behavior as above. 
A more spread-out initial distribution has the same qualitative behavior 
(not shown) except that it saturates early as expected. 
We note that the distributions for both Figs.(1,2) are {\em not} Gaussian 
during the meta-stable stage of the dynamics. Fig.~(3) shows results 
for $2\pi\hbar = 0.1$. Here the uncertainty principle constrains the initial 
$\rho$ to occupy a substantial portion of the available phase-space. We see 
again that for $\zeta \gg 1$, the quantum behavior is substantially different 
from the classical and there is almost no sensitivity to the environment 
(low entropy production). 
As $D$ increases, quantum-classical correspondence is indeed restored, 
although $D$ is now so large the dynamics are essentially that of the noise 
alone and $\dot{S_2}$ saturates at less than $\Lambda_2$. The abruptness of 
the tail in all our computations is most probably a numerical artifact 
due to the extremely small numbers being computed at that time; a detailed 
understanding of this tail is still absent. 
These results show that the entropy production rate for an arbitrary 
distribution in classically chaotic system saturates, after an exponentially 
rapid transient, to the sum of the positive generalized Lyapunov exponents 
of the system; further, the distribution settles to a meta-stable state. 
This is a clear signature of the underlying chaos, stable against perturbation. 
This behavior is echoed by quantum systems close to the classical limit, and 
the saturation is the precursor for quantum-classical correspondence in 
chaotic systems coupled to the environment. 

{\em Acknowledgement} - It is a pleasure to acknowledge discussions with
Bala Sundaram. 

\begin{figure}[htbp]
\caption{Time-dependence of the entropy $S_2(t)$ for the chaotic Cat Map 
for varying levels of coupling to the environment, as measured by $D$. 
Classical and quantal ($2\pi\hbar = 10^{-5}$) numerics were run for 
identical initial conditions and noise strengths $D$. The classical results 
are shown as points overlaid on the lines for the corresponding quantal data. 
There are two different initial conditions, one with high initial entropy 
$S_2(0)$ and the other with lower $S_2(0)$.  A reference line 
with slope equal to $\Lambda_2 = 0.9624$ is also shown.}
\end{figure}
\begin{figure}[htbp]
\caption{Quantal and classical results are shown with $2\pi\hbar =10^{-3}$,
a single initial condition and varying $D$. As $D$ is increased, the 
quantal behavior approaches that of the classical system. As in
Fig.~(1), a reference line with slope equal to $\Lambda_2 = 0.9624$ 
is also shown.}
\end{figure}
\begin{figure}[htbp]
\caption{As in Fig.~(2), but with $2\pi\hbar =10^{-1}$. The quantal system 
is almost unaffected by small $D$. The quantal behavior tends to the classical 
with increasing $D$; however, it is then dominated by noise and the slope is 
always less than that of the reference line shown with slope of 
$\Lambda_2 = 0.9624$.} 
\end{figure}

\begin{thebibliography}{99}

\bibitem{gaspard} 
P.~Gaspard, {\em Chaos, Scattering and Statistical Mechanics}
(Cambridge University Press, New York, 1998).

\bibitem{97_1} A.~K. Pattanayak and P.~Brumer, \pre {\bf 56}, 5174 (1997)
and references therein.

\bibitem{berman} G.~P.~Berman and G.~M.~Zaslavsky, Physica {\bf 91A},
450 (1978).

\bibitem{zp}W.~H.~Zurek and J.~P.~Paz, \prl {\bf 72}, 2508 (1994); 
Physica {\bf 83 D}, 300 (1995). 

\bibitem{entropy} These definitions are consistent with those in the
literature up to minus signs and constant factors. These forms are both 
specific cases of the generalized Renyi entropy\cite{schlogl} 
$S_{\beta} = \frac{1}{\beta -1}\ln({\rm Tr}[\rho^{\beta}])$.

\bibitem{schlogl}C.~Beck and F.~Schl\"ogl, {\em Thermodynamics of
chaotic systems}, (Cambridge University Press, N.Y., 1993).

\bibitem{97_2} A.~K. Pattanayak and P.~Brumer, \prl {\bf 79} 4131 (1997).

\bibitem{static}
See for instance I. Hamilton {\em et al}, \pra {\bf 25}, 3457 (1982);
K.~Takahashi, Prog. Theor. Phys. (Suppl.) {\bf 98}, 109 (1989);
Y.~Gu {\em et al}, Phys. Lett. A {\bf 229}, 208 (1997);
V.Latora {\em et al}, \prl {\bf 82} 520 (1999).

\bibitem{miller} P.~Miller and S.~Sarkar, \pre {\bf 58}, 4217 (1998);
 \pre {\bf 60}, 1542 (1999).

\bibitem{bag}B.~C.~Bag {\em et al}, Physica {\bf 125 D}, 47 (1999). 

\bibitem{quantum_cat} V. I. Arnold and A. Avez, {\em Ergodic Problems of 
Classical Mechanics} (Addison-Wesley, New York, 1989). The quantum 
version corresponds to J.~Wilkie and P.~Brumer, \pre {\bf 49}, 1968 (1994)
and A.~K. Pattanayak and P. Brumer, \prl {\bf 77}, 59, (1996).

\bibitem{decoh}
See W.~H. Zurek, Physics Today, {\bf 46}, 81 (October 1991) for a
popular review and extensive references therein.

\bibitem{footnote_n}
Since this is essentially independent of the size of $D$, it is in 
particular independent of $Dt$. This explains the results\cite{static} 
on the entropy production rates of classically chaotic systems subject to 
{\em static} coarse-graining. The averaging over the fine-scale structure 
in this case is not environmental and time-dependent but done by 
`binning' the distribution. 

\bibitem{expl}
This thus explains the result\cite{miller} that although the entropy 
production rate in a specific chaotic system qualitatively verified 
the conjecture in\cite{zp}, it did not quantitatively agree.

\bibitem{kosl}This is similar but not identical to the criteria
in ~\cite{zp} and also in A.~R.~Koslovsky, \prl {\bf 76}, 340 (1996).

\bibitem{decoh-corresp}S.~Habib {\em et al}, \prl {\bf 80}, 4361 (1998);
B. Mirbach {\em et al}, \prl {\bf 75}, 362 (1995).

\end{thebibliography}
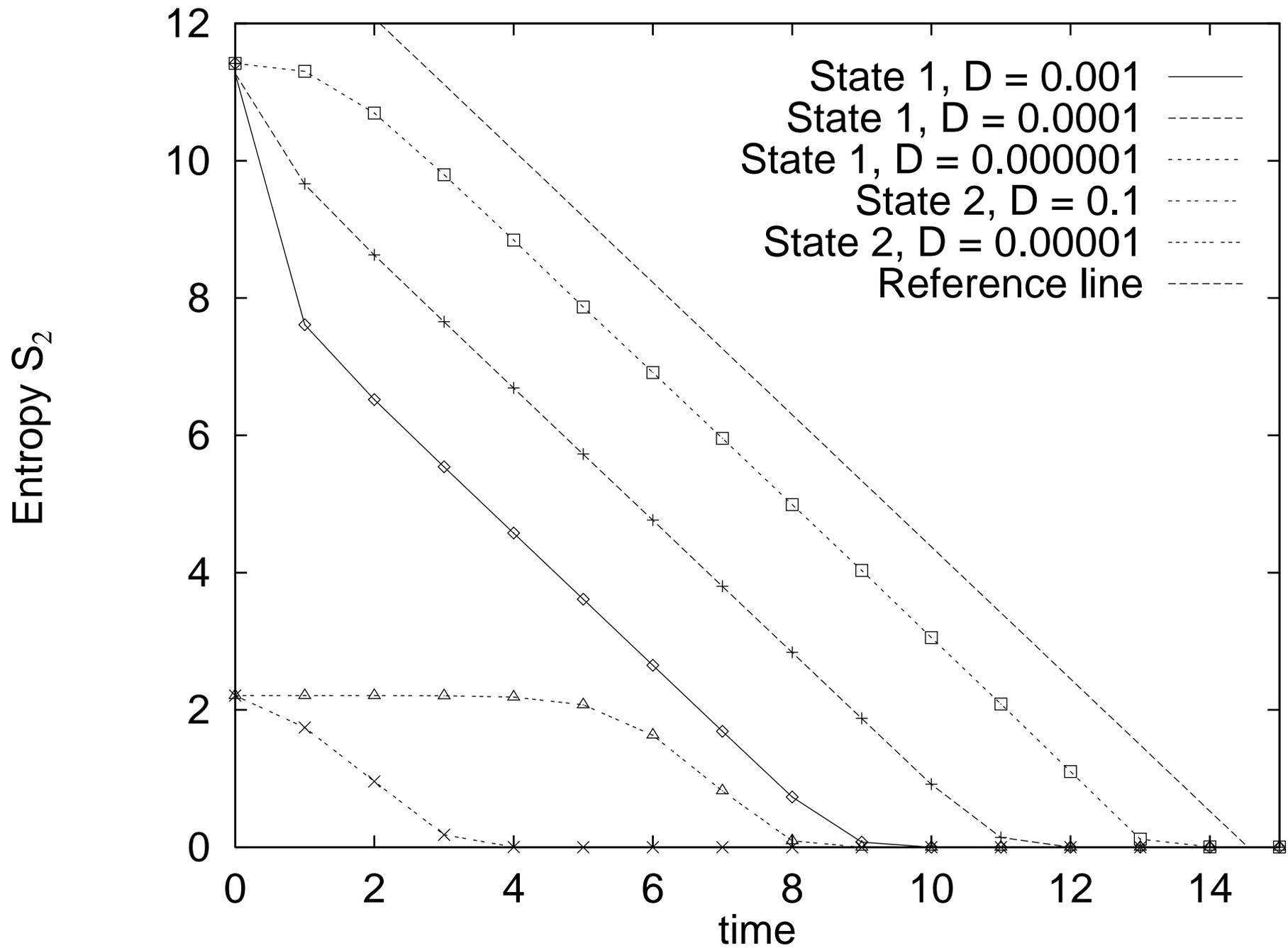
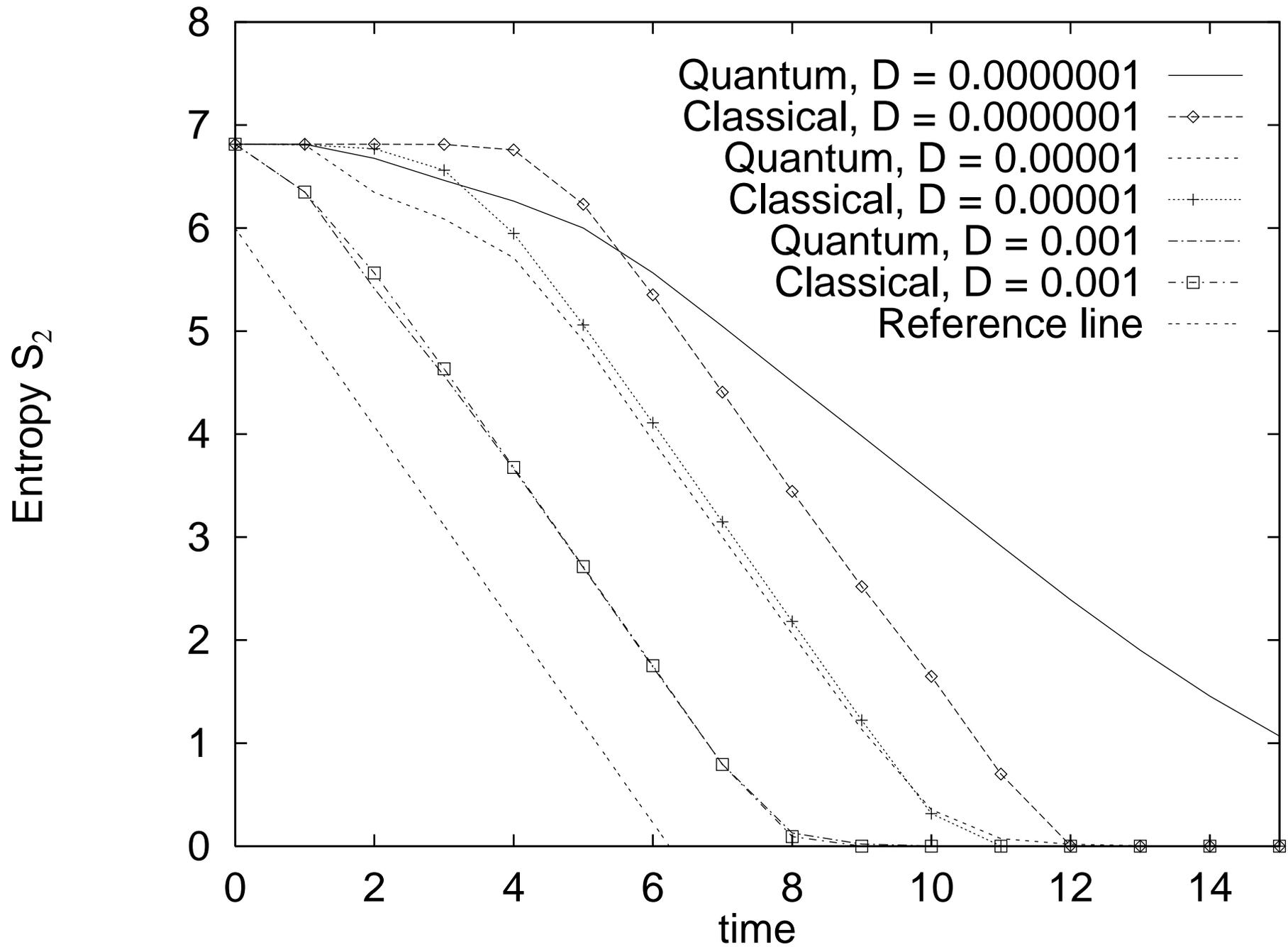
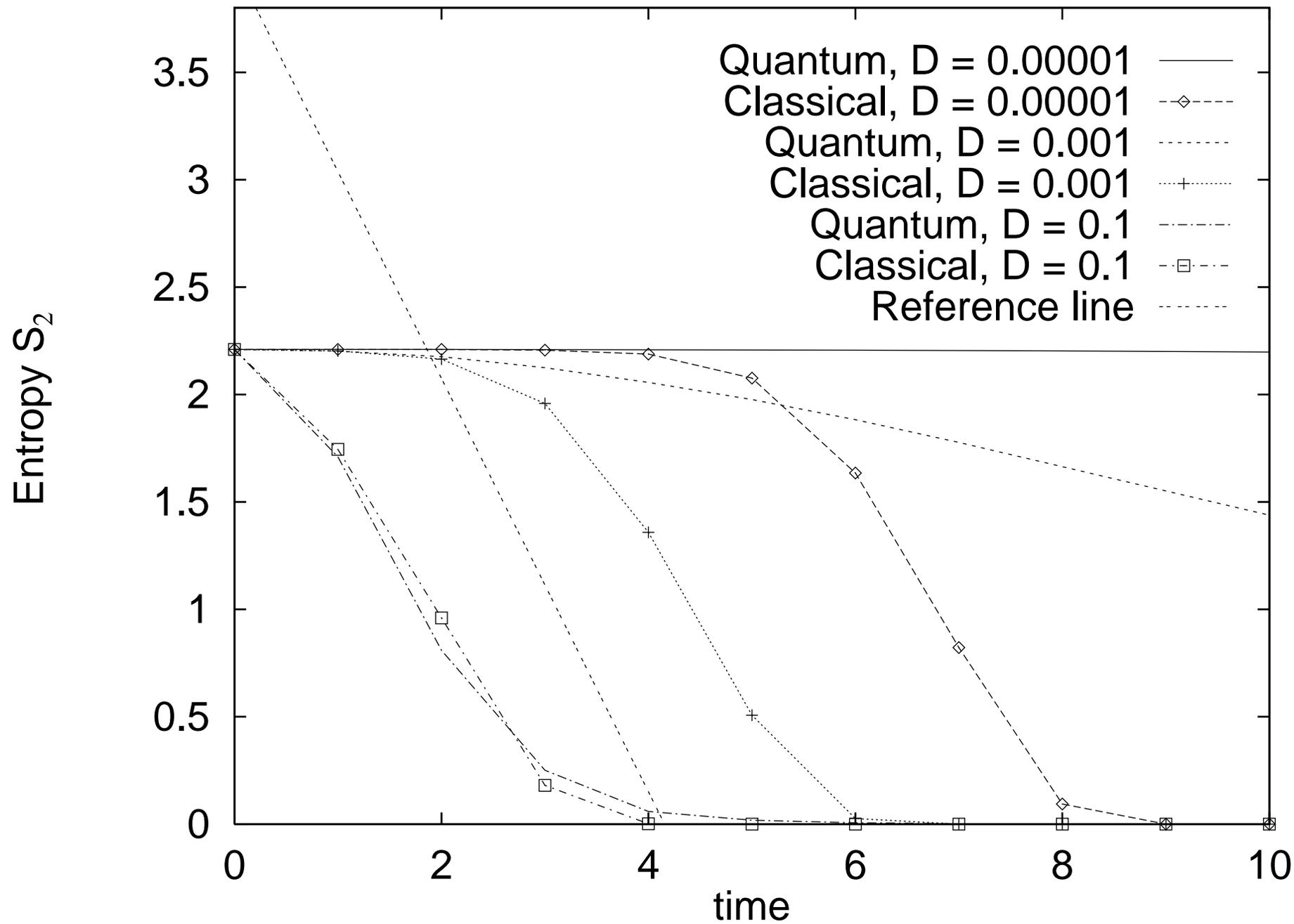
\end{document}